\documentclass[twocolumn, preprintnumbers,prl]{revtex4}
\usepackage{amsmath,amsfonts,amssymb,braket,graphicx}
\usepackage{setspace}
\usepackage[english]{babel}
\usepackage{graphicx}
\usepackage{dcolumn}
\usepackage{bm}
\usepackage{dsfont}
\usepackage{epstopdf}
\usepackage{gensymb}
\usepackage{nicefrac} 
\usepackage{color}

\usepackage{amsmath}
\usepackage{amsfonts}
\usepackage{amssymb}
\usepackage{amsthm}
\usepackage{makeidx}
\usepackage{graphicx}
\usepackage{bm}

\theoremstyle{remark}

\theoremstyle{definition}

\def\1{\mathchoice{\rm 1\mskip-4.2mu l}{\rm 1\mskip-4.2mu l}{\rm
		1\mskip-4.6mu l}{\rm 1\mskip-5.2mu l}}

\usepackage{amsmath}%
\usepackage{amsfonts}%
\usepackage{amssymb}%
\usepackage[cal=cm]{mathalfa}
\usepackage{graphicx}
\usepackage{braket}
\usepackage{sidecap}
\usepackage{color}
\usepackage{bbm}
\usepackage{nicefrac}
\usepackage{float}
\usepackage{placeins}
\usepackage[dvipsnames]{xcolor}
\usepackage[utf8]{inputenc}
\usepackage[normalem]{ulem}
\usepackage{nicefrac}
\newcommand{\stkout}[1]{\ifmmode\text{\sout{\ensuremath{#1}}}\else\sout{#1}\fi}
\usepackage{dsfont}
\usepackage{xcolor, soul} 
\usepackage[normalem]{ulem} 
\usepackage{siunitx}
\usepackage[citebordercolor=blue]{hyperref}

\DeclareSIUnit\gauss{G}

\newcommand{\angstrom}{\textup{\AA}}

\begin{document}
	
	\title{Spin - Rotation Coupling Observed in Neutron Interferometry}
	\author{Armin Danner$^1$}
	\author{B\"ulent Demirel$^1$}
	\author{Wenzel Kersten$^1$}
	\author{Richard Wagner$^1$}
	\author{Hartmut Lemmel$^{1,2}$}
	\author{Stephan Sponar$^1$}
	\author{Yuji Hasegawa$^{1,3}$}
	\email{yuji.hasegawa@tuwien.ac.at}
	\affiliation{%
		$^1$Atominstitut, TU Wien, Stadionallee 2, 1020 Vienna, Austria \\
		$^2$Institut Laue Langevin, 38000 Grenoble, France\\
		$^3$Department of Applied Physics, Hokkaido University, Kita-ku, Sapporo 060-8628, Japan}
	
	\date{\today}
	
	\hyphenpenalty=800\relax
	\exhyphenpenalty=800\relax
	\sloppy
	\setlength{\parindent}{0pt}
	
	\noindent
	
\begin{abstract}
		%
		Einstein's theory of general relativity and quantum theory form the two major pillars of modern physics. However, certain inertial properties of a particle's intrinsic spin are inconspicuous while the inertial properties of mass are well known. Here, by performing a neutron interferometric experiment, we observe phase shifts arising as a consequence of the spin's coupling with the angular velocity of a rotating magnetic field. The resulting phase shifts linearly depend on the frequency of the rotation of the magnetic field. Our results agree well with the predictions derived from the Pauli - Schr\"odinger equation. 
	\end{abstract}
	
	\maketitle
	
\emph{Introduction.---}%
The principle of equivalence of inertial and gravitational masses is a corner stone of Einstein's theory of general relativity \cite{EinsteinBook}. It follows from this principle that one cannot locally distinguish between inertial forces and pseudo-forces. Examples of pseudo-forces are the gravitational force, as experienced in the presence of a massive object, or Coriolis and centrifugal forces, which originate from circular motion of an observer in a non-inertial frame of reference. In terms of wave phenomena and in a rotating frame, the respective phase shifts are described by additional couplings. The Sagnac effect \cite{Sagnac13} refers to the observed phase shift induced between two counter rotating light waves in a rotating interferometer. This phase shift is proportional to the scalar product of the rotation frequency and the area of the installed interferometer. This can also be written as a coupling $\sim\boldsymbol\Omega\cdot\boldsymbol L$ between the rotation vector $\boldsymbol\Omega$ and the orbital angular momentum $\boldsymbol L$ of the light wave around the center of rotation. 
	\par 
In quantum theory the inertial properties of a particle are influenced not only by its inertial mass, but also by its spin. When solving Dirac's equation in accelerated frames of reference in the non-relativistic regime, the Hamiltonian of a particle includes the term $-\boldsymbol\Omega\cdot\boldsymbol J$ \cite{Hehl90}, where $\boldsymbol\Omega$ is the rotation vector of the frame and $\boldsymbol {J=L+S}$ is the total angular momentum of the particle with the contribution $\boldsymbol S$ of the spin angular momentum. The contribution $-\boldsymbol\Omega\cdot\boldsymbol L$ of the Sagnac effect was first demonstrated experimentally for the de Broglie waves of neutrons in the late 1970s \cite{Werner79}. 
	\par 
	To measure the spin contribution, Mashhoon first published a proposal by S. A. Werner for an experiment involving a rotating neutron interferometer \cite{Mashhoon88} (in an arrangement insensitive to the Sagnac and gravity effects). In the further course, Mashhoon \textit{et al.} suggested interferometer setups where longitudinally polarized neutrons pass through a slowly rotating spin flipper \cite{Mashhoon98} which is in turn equivalent to a rotating magnetic field \cite{Mashhoon06}. The authors of \cite{Mashhoon98} stated that "the phenomenon of spin-rotation coupling is of basic interest since it reveals the inertial properties of intrinsic spin." For further theoretical contributions about spin-rotation coupling consider \cite{Mashhoon1995, Soares1996, Ryder98, Ryder2008, Arminjon2014}. Recently, we reported on neutron polarimeter experiments \cite{Demirel15, Danner19} whose measurement results can be attributed to the coupling of the neutron's spin with the rotation of a magnetic field. However, the results of this experiments rely on the rotation of the spin orientation which can also be described by the semi-classical Bloch equations. 
	\par 
	In this letter, we present the results of the neutron interferometric experiment as suggested by Mashhoon~\cite{Mashhoon06}. The relative phase between the partial wave functions of paths I and II in the interferometer is directly measured. By applying a direct measurement of the relative phase, instead of measuring the rotation of the spin vector as in a polarimeter experiment, the purely quantum mechanical aspect of the spin-rotation coupling is demonstrated.
	
	Neutron interferometry \cite{RauchBook} is a technique to observe the interference effect of matter waves passing through a perfect silicon-crystal interferometer. It is an established, powerful tool to investigate fundamental quantum mechanical concepts with massive particles \cite{Rauch74}. Using neutron interferometry the $4\pi$ spinor-symmetry of fermions \cite{Rauch1975, Werner1975}, the spin-superposition law \cite{Badurek1983, Summhammer1983} and the equivalence principle \cite{Colella1975, Bonse1983} have been demonstrated.
	
	\emph{Theory.---}%
	Let us consider an observer rotating relative to an inertial observer. The wave function~$\psi'(\boldsymbol r,t)$ with respect to the rotating frame of reference, is given by the wave function $\psi(\boldsymbol r,t)$ in the inertial frame as $\psi'=\hat U\psi$. The unitary operator $\hat U$ is given by $\hat U={\mathrm{exp}}(\mathrm i\,\boldsymbol\Omega\cdot\boldsymbol J t/\hbar)$, with $\boldsymbol J$ being the total angular momentum, consisting of orbital and spin angular momentum. If the wave function $\psi$ satisfies the Schr\"odinger equation $\hat H\psi={\mathrm i}\hbar\, \partial \psi/\partial t$, the wave function $\psi'$ represents a solution of the Schr\"odinger equation $\hat H'\psi'={\mathrm i}\hbar\, \partial \psi'/\partial t'$ with $\hat H'=\hat U \hat H \hat U^\dagger-\gamma \,{\boldsymbol\Omega\cdot\boldsymbol J}$, with the Lorentz factor $\gamma$. A detailed comparison of the latter equations \cite{Hehl90, Ryder98} reveals the existence of a new effect associated with the coupling of intrinsic spin with rotation which is expressed by the Hamiltonian $\delta\hat H^\prime_{\mathrm{SR}}=-\gamma\,\boldsymbol\Omega\cdot\boldsymbol S$.
	
	As suggested by Mashhoon, the effect can indeed be derived as done before by solving the Pauli --  Schr\"odinger equation in the lab frame for the interaction of the spin of a free neutron in a magnetic field with angular velocity $\Omega$. 
	
	For a neutron propagating in $+y$-direction through an uniformly rotating magnetic field, which is expressed as $\boldsymbol B(\Omega, t) = B_1\left(\cos\left (\Omega t\right), 0, \sin \left(\Omega t\right)\right)^T$, a solution is given by 
	\begin{equation}
	\psi(y,t)=\big(\frac{1}{\sqrt{2\pi}}e^{{\mathrm i}k y}\big)\big(e^{-{\mathrm i}\frac{\hbar k^2}{2m}t}\big)\xi(t), 
	\end{equation}
	where $\xi(t)$ generates the rotation of the initial spin state in the rotating frame $\xi_{\mathrm{rot}}(0)$ and is given as 
	\begin{equation}
	\xi(t)=e^{\frac{\mathrm i}{\hbar}\Omega S_y t}e^{-\frac{\mathrm i}{2}\boldsymbol\alpha_{\mathrm{rot}}\cdot\boldsymbol \sigma_{\mathrm{rot}}}\xi_{\mathrm{rot}}(0)=\hat U(\Omega)\hat U(\boldsymbol\alpha_{\mathrm{rot}})\xi_{\mathrm{rot}}(0)
	\end{equation}
	with the vector $\boldsymbol\sigma_{\mathrm{rot}}$ comprising the Pauli matrices. The operator $e^{\frac{\mathrm i}{\hbar}\Omega S_y t}$ is the transformation from the rotating into the laboratory frame while the operator $e^{-\frac{\mathrm i}{2}\boldsymbol\alpha_{\mathrm{rot}}\cdot\boldsymbol \sigma_{\mathrm{rot}}}$ describes the spin evolution in the rotating frame with the operators acting in the  respective frame. The magnitude of the rotation vector $\boldsymbol \alpha_{\mathrm{rot}}(t)=(\omega_1t,\Omega t,0)^T $ is given by 
	\begin{equation}
	\label{equ3}
	\alpha(t)=t\sqrt{\omega_1^2+\Omega^2}, 
	\end{equation}
	where the definition of the Larmor frequency $ \omega_1=-\frac{2\mu}{\hbar}B_1 $ is used. Both operators $\hat U(\Omega)$ and $\hat U(\boldsymbol\alpha_{\mathrm{rot}})$ include a term $\sim\boldsymbol \Omega\cdot\boldsymbol S$ but the operator $\hat U(\boldsymbol\alpha_{\mathrm{rot}})$ also includes the effect of the Larmor precession.

	\emph{Experimental Set-Up.---}%
	\begin{figure}[!b]
		\includegraphics[width=0.48\textwidth]{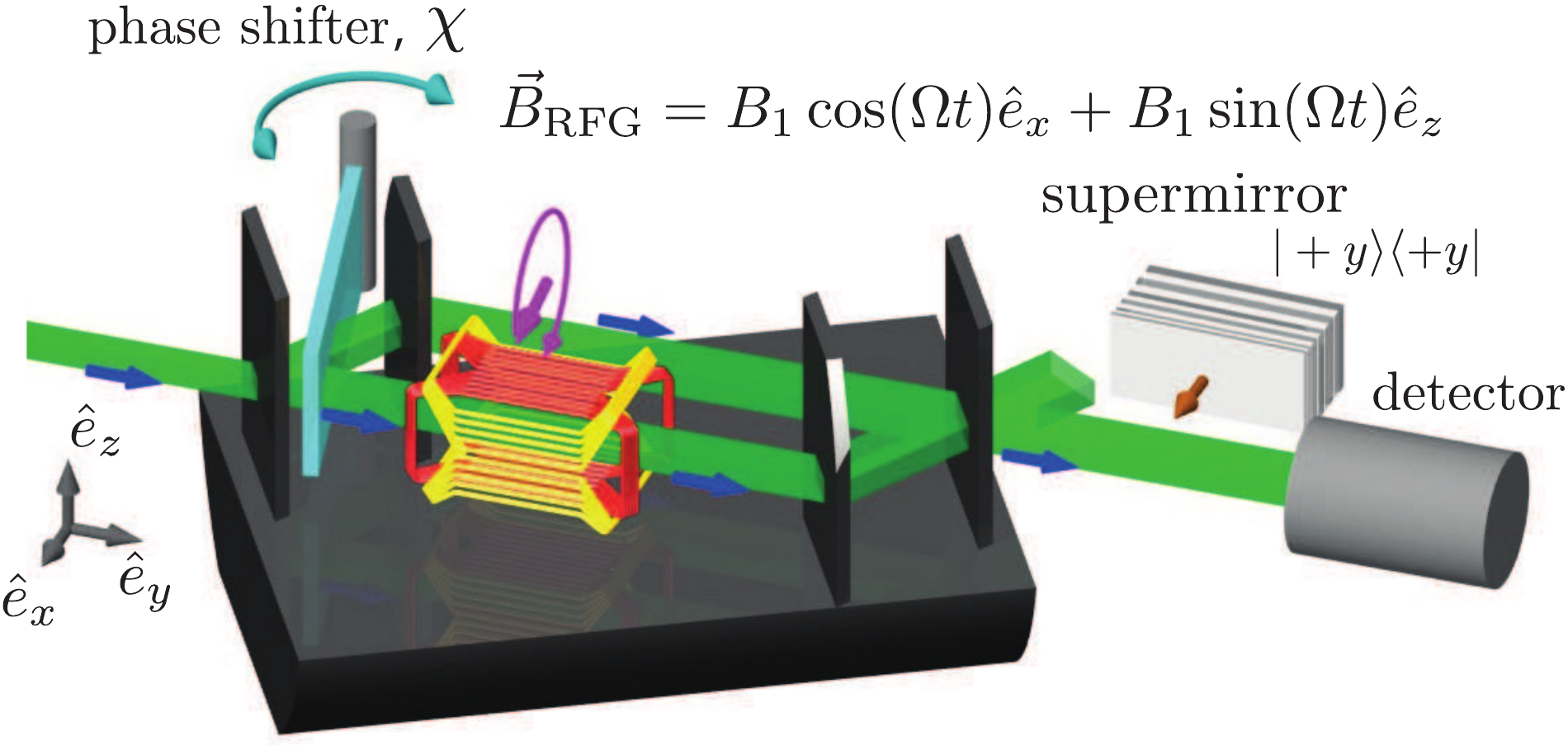}
		\caption{Monochromatized neutrons, with polarization parallel to the direction of propagation ($\ket{+y}$), enter the interferometer. Inside the interferometer the rotating field generator (RFG) creates a magnetic field $\boldsymbol B_{\rm{RFG}}(\Omega,t)$ in path $\ket{II}$, rotating in a plane perpendicular to the neutron beam. After recombination at the last interferometer plate, the neutrons in foreward direction are detected in a He-3 counter tube. Spin directions are indicated with blue arrows, the magnetic field in pink. The phase shifter plate~($\chi$) consists of a slab of sapphire and is rotated to record interferograms. \label{fig:setup}}
	\end{figure}
	To test the existence of spin-rotation coupling the following experiment was carried out at the neutron interferometry station S18 at the High-flux reactor of the Institut Laue-Langevin in Grenoble. The set-up after the monochromator and polarizer is schematically illustrated in Fig.\,\ref{fig:setup}. The incoming beam, with initial polarization parallel to the direction of flight ($\hat e_y$), is coherently split at the first interferometer plate in two parts, the transmitted beam $\ket{I}$ and the reflected beam $\ket{II}$. Both beams are reflected by plates three and two, respectively, and recombined at the fourth plate. The intensity of the The +y-component of the spin orientation of the exit beam in foreward direction (O-beam) can be analyzed with a CoTi supermirror array, from now on called \textit{supermirror}. The intensity transmitted by the supermirror is measured by the detector. The beam in the diffracted direction (H-beam) is not used in the present experiment. In the interferometer a coil, henceforth called \emph{rotating field generator} (RFG) \cite{Danner19}, is placed in path II. The field of the RFG is generated by a double coil arrangement for the $x$- and $z$-direction which can be used to implement a static field or a field rotating about the beam direction with oscillating electric currents which are phase shifted by $\pi/2$, resulting in a magnetic field which rotates in time. The coil in the interferometer is water-cooled to stabilize the phase relation between both paths. The coil's geometry is optimized to have no wire in the direct beam path \cite{Geppert14} thereby ensuring a high interferometric contrast. The phase shifter plate is used to induce a relative phase shift $\Delta\chi$ between paths $\ket{I}$ and $\ket{II}$. 
	\par
	
	The RFG generates a rotating magnetic field $\boldsymbol B_\mathrm{RFG}$ around the $y$-axis. The incoming spin is parallel to the field rotation vector $\boldsymbol\Omega$, i.e., orthogonal to the rotating magnetic field $\boldsymbol B_\mathrm{RFG}$, when entering the RFG. 
	\par
	When the incident spin state is $\xi(0) = \ket{+y}$ and $\alpha(t_1) = 2\pi$, with the time $t_1$ it takes for the neutron to fly through the region of the rotating magnetic field, the spin state after a cyclic evolution on the Bloch sphere becomes $\xi(t_1)= - e^{i\Omega t_1/2} \ket{+y}$ which is solely dependent on the henceforth called \emph{Mashhoon Phase} $\Omega t_1/2$.
	\par
	To adjust the case of $\alpha(t_1) = 2\pi$ of cyclic evolution paths of the spin orientation in the RFG, the amplitude $B_1$ of the rotating field is scanned. The cyclic paths are generated at a certain amplitude for each frequency $f=\Omega/2\pi$, from \SIrange{0}{20}{\kilo\hertz}. The necessary amplitudes decreased with increasing frequency as expected through (\ref{equ3}).
	\par 
	As path $\ket{I}$ will serve as a reference for the phase induced in path $\ket{II}$, the spin orientations of both paths must be parallel at the last plate of the interferometer to guarantee maximum interference. At each phase shifter orientation, the counts per 20 seconds are recorded for different frequencies with the appropriate sinusoidal currents in the RFG. In the static case, $\Omega = 0$, the neutron spin rotates around the static B field in $x$-direction by an angle of $2\pi$, hence returning to its initial polarization. 
	

	\begin{figure}[!b]
		\includegraphics[width=0.35\textwidth]{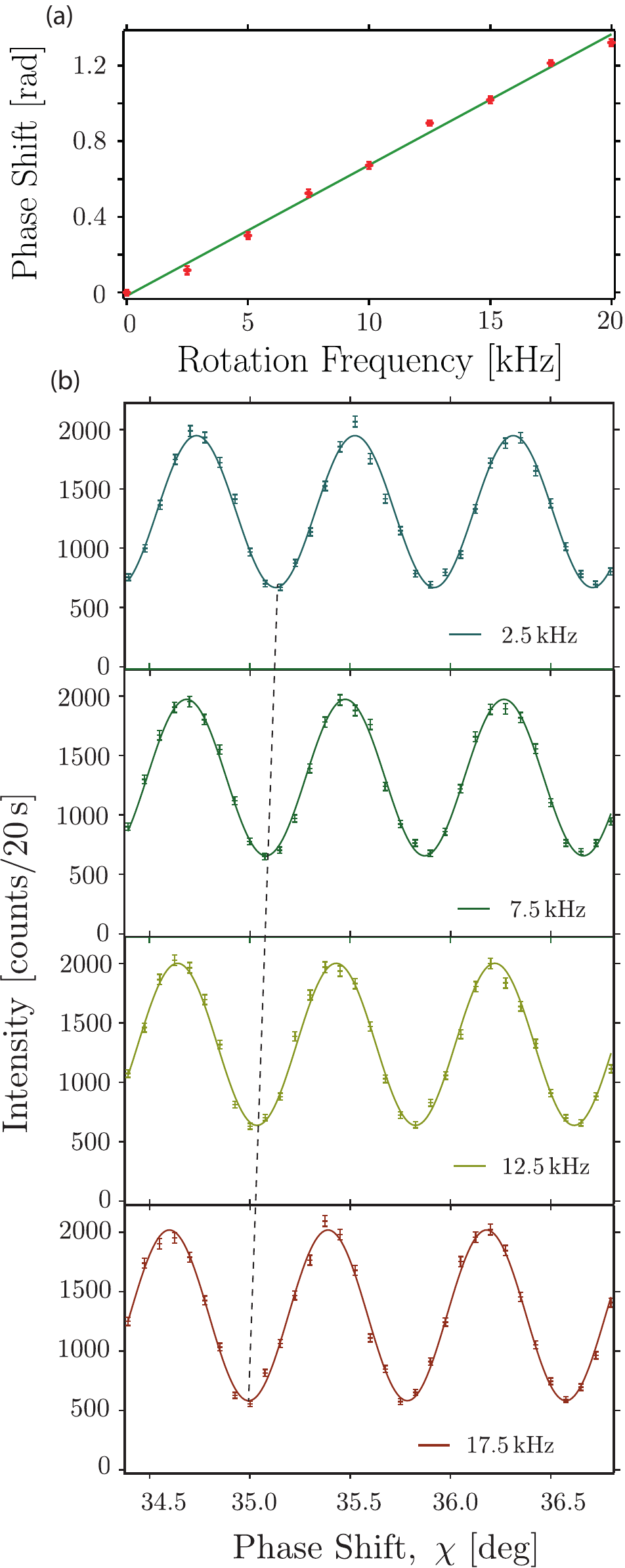}
		\caption{(b) A subset of interferograms recorded as intensity oscillations by rotating the phase shifter orientation, together with respective sinusoidal fit functions, for 2.5\,kHz, 7.5\,kHz, 12.5\,kHz and 17.5\,kHz. With increasing frequency the interferograms are also continuously shifted. Error bars indicate $\pm$ one standard deviation. Gray dotted line is to guide the eye for the observed phase shift. (a) Linearly fitted phase of interferograms relative to the static field dependent on the rotation frequency of the magnetic field inside the RFG. 
		}
		\label{fig:interferogram}
	\end{figure}
	
	\emph{Results.---}%
	This way the interferograms in Fig.\,\ref{fig:interferogram}(b) were recorded by changing the phase shifter orientation -- each interferogram with a different frequency of the rotating magnetic field in the RFG. The interferograms are continuously shifted with increasing frequency. The fitted phases of the interferograms relative to the static case are plotted in Fig.\,\ref{fig:interferogram}(a). The expected phase $\Omega t_1/2$ is linearly dependent on the rotation frequency $f$. The measurement results comply with this linear behaviour. The deviations from the according linear fit are systematic due to misadjustment of the $x$ and $z$-amplitudes of the rotating magnetic field inside the RFG. 
	\par
	\begin{figure}[!b]
		\includegraphics[width=0.35\textwidth]{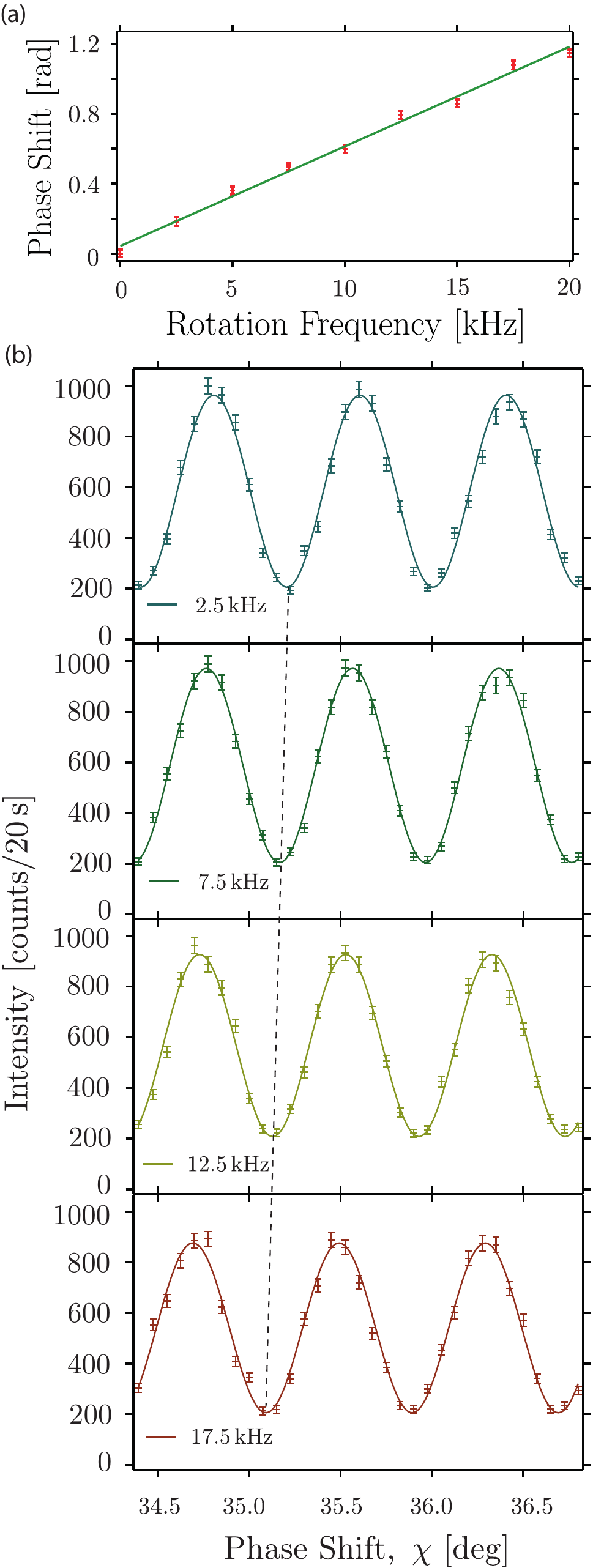}
		\caption{(b) A subset of interferograms recorded as intensity oscillations by rotating the phase shifter orientation, together with respective sinusoidal fit functions, for 2.5\,kHz, 7.5\,kHz, 12.5\,kHz and 17.5\,kHz. With increasing frequency the interferograms are also continuously shifted. Error bars indicate $\pm$ one standard deviation. Gray dotted line is to guide the eye for the observed phase shift. (a) Linearly fitted phase of interferograms relative to the static field dependent on the rotation frequency of the magnetic field inside the RFG. 
		}
		\label{fig:interferogramSM}
	\end{figure}
	
	At this point we want to emphasize the pure quantum nature of spin-rotation coupling, i.e., that it cannot be explained semi-classically in terms of rotations of the polarization vector. To demonstrate this, the experiment is also conducted with the supermirror inserted in the O beam (Fig.\,\ref{fig:interferogramSM}). This guarantees that the final polarization vector is unchanged. The results are reproduced evidently (only a lower count rate is observed because of the additional neutron optical components).

	\emph{Conclusion and Discussion.---}%
	A neutron interferometer experiment was carried out to test the theoretical prediction of a spin-rotation coupling. The coupling in our experiment is established between the rotation of a magnetic field and the spin of the neutron. The results confirm the existence of the spin-rotation coupling and its linear dependency on the rotation frequency. As the amplitude of the rotating field decreased to adjust the case of $\alpha(t_1)=2\pi$, the phase shift in the interferograms is not due to a Zeeman effect. In contrast to previous neutron polarimeter experiments \cite{Demirel15, Danner19}, the phase shift is a purely quantum mechanical effect and cannot be explained by the semi-classical Bloch equations. 
	\FloatBarrier
	\emph{Acknowledgements.---}%
	We are grateful to Mario Pitschmann for helpful discussion. This work was financed by the Austrian Science Fund (FWF), Projects No. P27666-N20 and P30677-N20.

\bibliography{Bibliography}
 
\onecolumngrid
 
\appendix
 
\section{Supplementary Material}

In this Supplemental Material we present further technical details of the experimental setup, the adjustment procedure, as well as additional measurement data.
\begin{figure*}[!b]
	\includegraphics[width=0.7\textwidth]{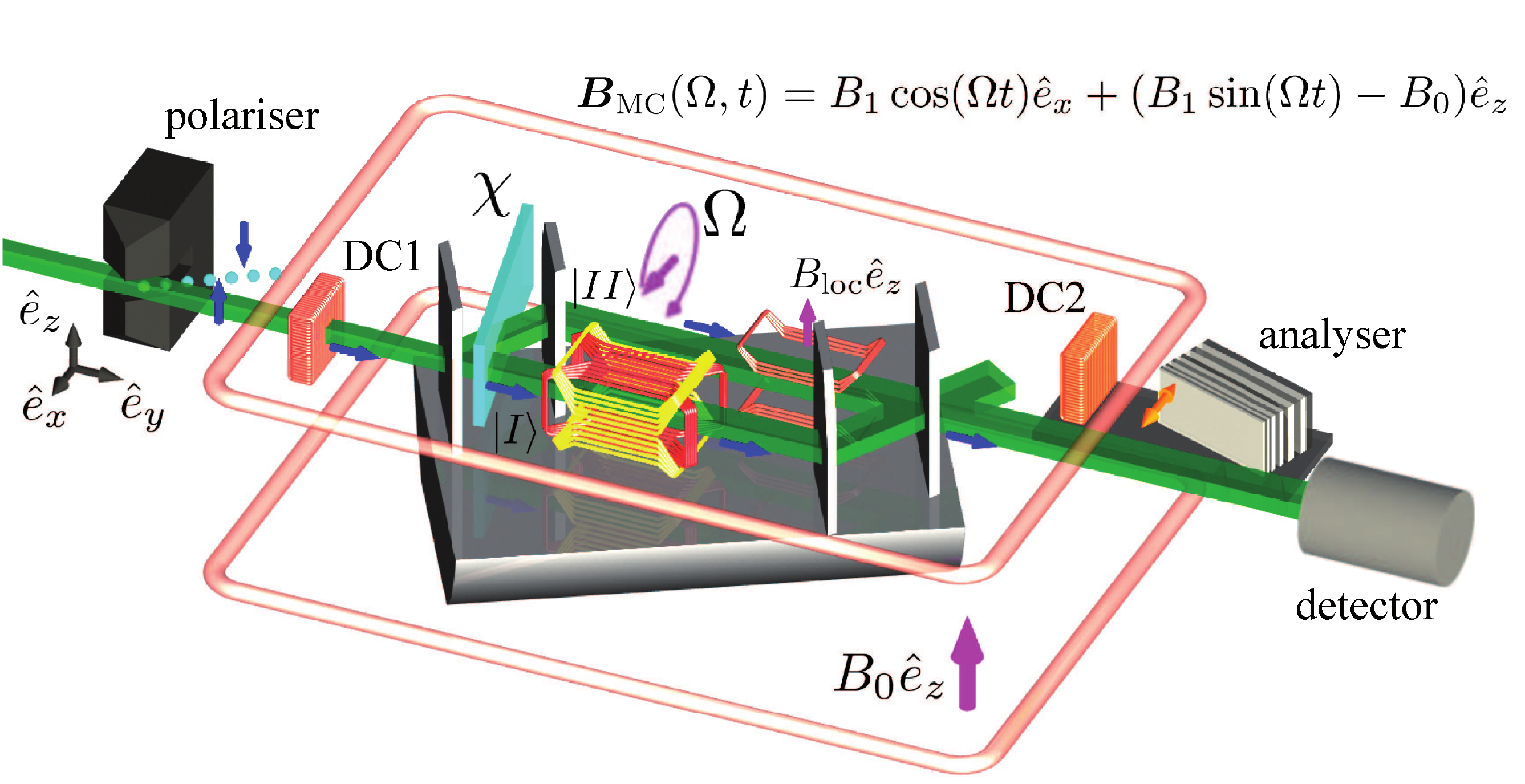}
	\caption{Monochromatized neutrons pass through a polarizing magnet (polarizer), selecting only the $\ket{+z}$-spin component to fulfill the Bragg-condition inside the interferometer. Direct-current spin-rotator DC1 turns the spin in the direction of flight. Inside the interferometer, the RFG creates a magnetic field $\boldsymbol B_{\rm{RFG}}(\Omega,t)$ in path $\ket{I}$, rotating with angular velocity $\Omega$ in a plane perpendicular to the neutron beam. In path $\ket{II}$, a local z-field $B_{\rm{loc}}$ is mainly used to compensate different times of interaction with the guide-field $B_0$ between the paths because of the length of the RFG. In this process, deviations of the guide field in both paths are also compensated. After recombination at the last interferometer plate, the neutrons in foreward direction are detected in a He-3 counter tube. For magnetic adjustment DC2 and the analyzer supermirror (analyzer) can be placed in front of the detector but are removed before the final measurement. Spin directions are indicated with blue arrows, magnetic fields in pink. The phase shifter plate ($\chi$) consists of a slab of sapphire and is rotated to record interferograms. \label{fig:setupSupp}}
\end{figure*}

\subsection{Experimental Setup}

The experimental setup applied for the measurement of spin-rotation coupling in neutron interferometry is schematically illustrated in Fig.\,\ref{fig:setupSupp}.  A beam of monochromatized neutrons with a wavelength $\lambda=\SI{1.9}{\angstrom}$ is split by a magnetic prism into two divergent, antiparallelly polarized sub-beams with spin states $\ket{\pm z}$. Next the neutrons enter a static magnetic guide field region, consisting of a coil in Helmholtz configuration that is placed between polarizer and analyzer and generates a field $\boldsymbol B_0=9\,{\rm G}\,\hat{\boldsymbol e}_z$. The purpose of the guide field is to prevent depolarization because of magnetic stray fields, thereby upholding the degree of polarization of $P\sim 0.99$. Inside the guide field, a direct current spin rotator (DC1) is placed in front of the interferometer which rotates the polarization vector by $\pi/2$ from $\pm z$ to $\pm y$, the flight direction of the neutrons. They traverse to a silicon perfect crystal interferometer which is aligned such that only the $+y$ polarized sub-beam fulfills the Bragg condition of the lattice at the first plate of the interferometer. Consequently, the $-y$ polarized sub-beam never reaches the detectors (in practice it is absorbed by a Cd-slab, which is not depicted in Fig.\,\ref{fig:setupSupp}). 

In path $\ket{I}$ of the interferometer, the RFG generates a rotating magnetic field denoted as $\boldsymbol B_{\rm{RFG}}(\Omega,t)=B_1\cos(\Omega t)\hat e_x+(B_1\sin(\Omega t)-B_0)\hat e_z$, rotating with angular velocity $\Omega$ and amplitude $B_1$ in a plane perpendicular to the neutron beam. This coil is the key element in the neutron optical setup and induces the spin-rotation coupling that shall be observed in the actual experiment. In path $\ket{II}$, a Larmor accelerator \cite{Geppert14} produces a local z-field $B_{\rm{loc}}$. This field is used to compensate differences in the Larmor precession angles (induced by the guide-field $B_0$) acquired the two paths. 
Finally a direct current spin rotator (DC2) as well as a spin analyzer, which only transmits $\ket{+z}$-spins, are placed in front of the detector for the adjustment procedure which is described below. 

\subsection{Adjustment Procedure}

\begin{figure*}[!h]
	\includegraphics[width=0.5\textwidth]{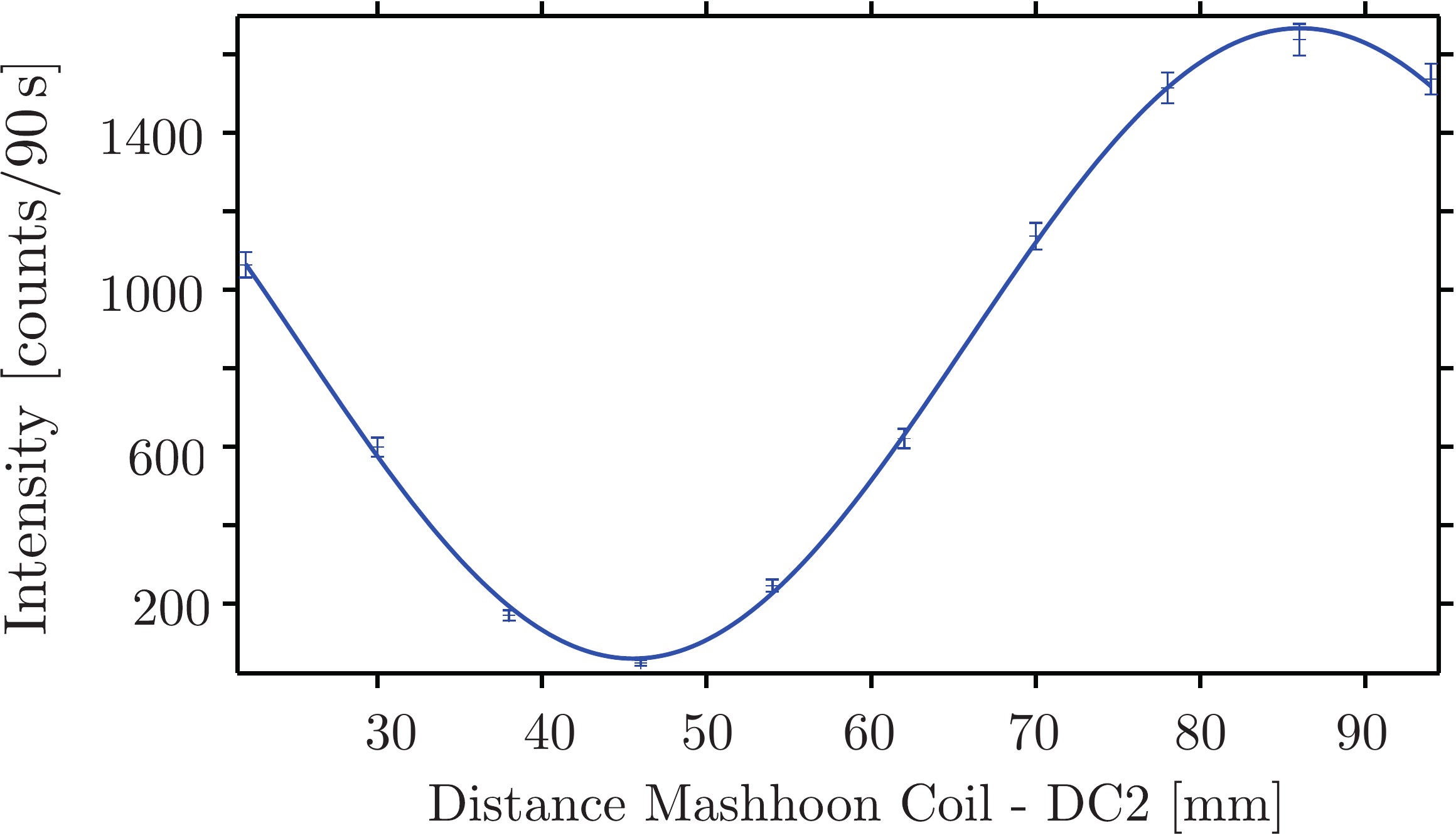}
	\caption{Intensity modulation due to variation of the distance between DC 1 and RFG. Both devices are  set to induce a $\pi/2$ spin rotation. Error bars indicate $\pm$ one standard deviation.\label{fig:Dist}}
\end{figure*}

The coil DC1 is used to prepare the spin state $\ket{+y}$ for the RFG while DC2 and analyzer are used in the adjustment procedure of the experiment to analyze the $\ket{+y}$ spin component. As Larmor precession is induced inside the guide field between the coils, their distances need to be adjusted such that the input for the RFG and DC2 is a $\ket{+y}$ spin state. To do this, path $\ket{II}$ is blocked with a Cadmium absorber, which is not depicted in Fig\,\,\ref{fig:setupSupp}. Coil DC1 rotates the spin by $\pi/2$ into the $x$-$y$-plane by Larmor precession about the static magnetic field inside the coil, which is pointing in $+x$-direction. Another rotation of $\pi/2$ about the same axis inside the RFG causes a minimum intensity in the case of a Larmor precession with an angle $2\pi$ in the guide field in between DC1 and RFG with Larmor frequency $\omega_0=-\frac{2\mu}{\hbar} B_0$. The observed intensity oscillation is depicted in Fig.\,\ref{fig:Dist}. The positions of DC1 and the RFG are fixed at the minimum intensity. The same procedure is repeated to adjust the position of DC2 relative to the RFG. 
\begin{figure}[!b]
	\includegraphics[width=0.81\textwidth]{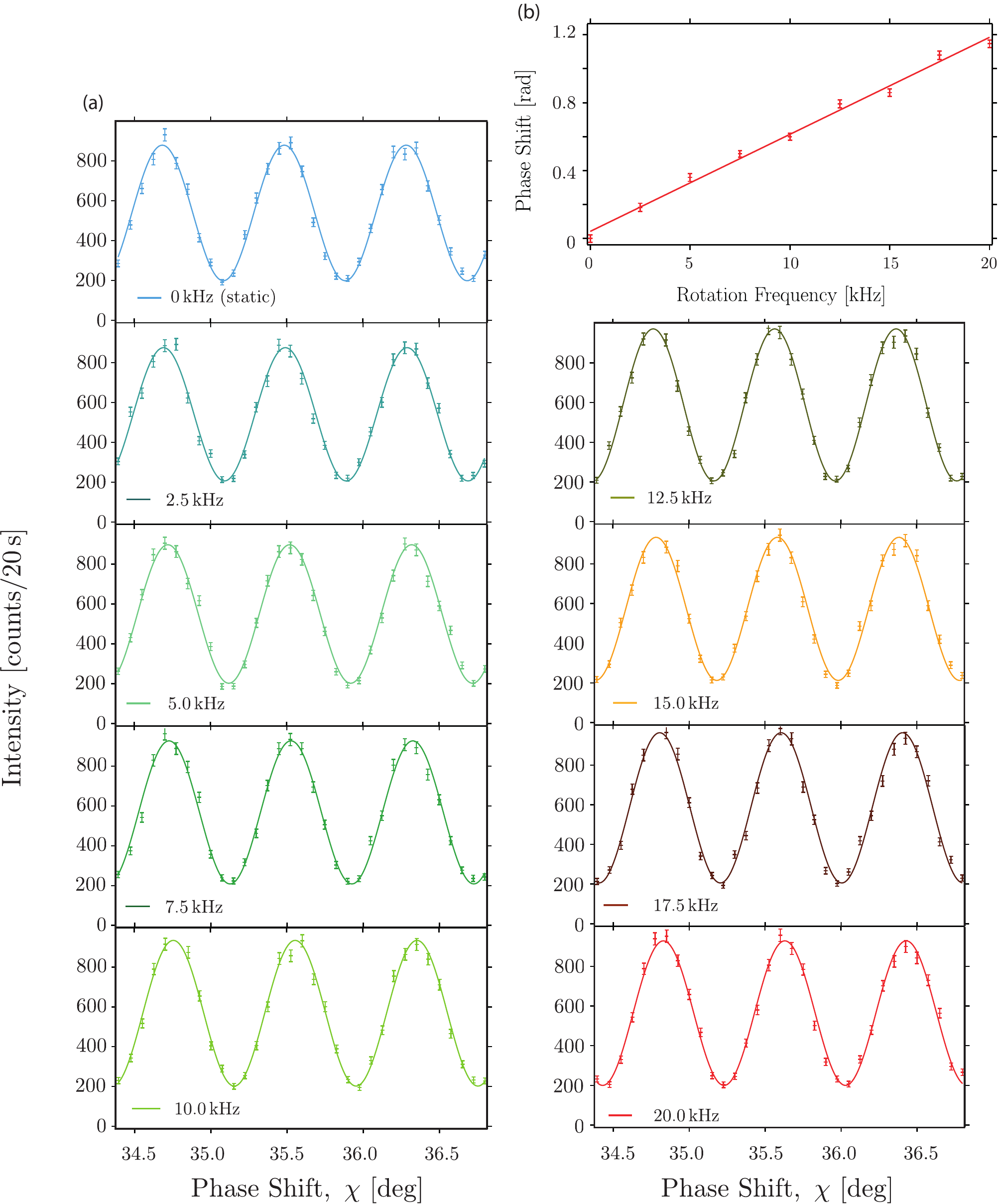}
	\caption{All interferograms recorded as intensity oscillations by rotating the phase shifter orientation with Supermirror and DC 2 inserted. With increasing frequency the interferograms are again continuously shifted. 
	}
	\label{fig:interferogramSM}
\end{figure}
\begin{figure}[!b]
	\includegraphics[width=0.8\textwidth]{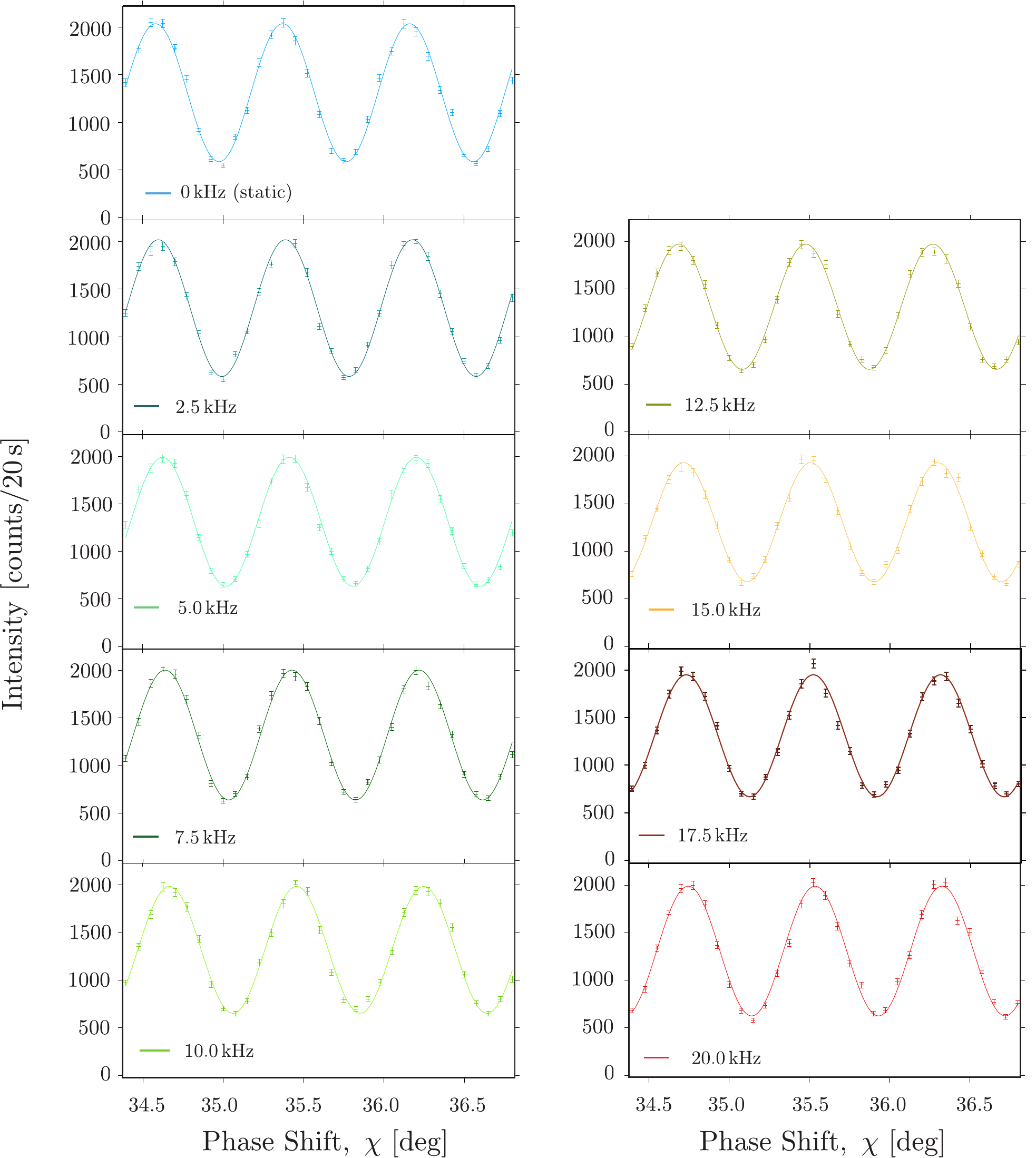}
	\caption{All interferograms recorded as intensity oscillations by rotating the phase shifter orientation. With increasing frequency the interferograms are also continuously shifted. 
	}
	\label{fig:interferogram}
\end{figure}

%

The RFG is supplied from now on with oscillating and $\pi/2$ phase shifted currents to produce the rotating field while each DC coil produces a $\pi/2$ rotation. For each frequency (\SIrange{0}{20}{\kilo\hertz}), the amplitudes of both currents through the RFG are simultaneously increased from zero and its resulting minimum intensity until the same minimum intensity is produced again. This signifies the case of a cyclic rotation of the polarization vector inside the Mashhoon box.
\par
As we want to investigate a Mashhoon phase shift in the interference between otherwise identical wave functions, both paths need to be recombined with aligned spin states. Therefore, the $z$-coil $B_{\mathrm{loc}}$ in path II is used, serving as Larmor accelerator. This coil is required since inside the RFG the $z$-field offset locally compensates the guide field, such that there is no static guide field present inside the RFG. 
To adjust the case of spin alignment, the absorber (not depicted) is switched to block path $\ket{I}$ and the $\pi/2$-rotations by of both DC1 and DC2 are maintained. The current in the $z$-coil for the magnetic field $B_{\rm{loc}}$ in path $\ket{II}$ is scanned. The spin orientations of both beams are parallelly adjusted at the last plate with that current for the Larmor accelerator which produces a minimum intensity.
The above setup procedure is necessary to adjust the currents and relative positions of DC1, DC2, the Larmor accelerator and the RFG. The absorber is removed for the following measurements.  
\par 
Interferograms were produced with the described setup. The observed results are plotted in Fig.\,\ref{fig:interferogramSM}, showing a phase shift with spin analysis.   

Unlike described in the main text, only the last step of the measurements were conducted with absorber, DC2 and analyzer removed. All obtained interferograms, contributing to the plot of the final results (main text Fig.\,3) are plotted in Fig.\,\ref{fig:interferogram}. The only difference is a higher count rate due to the less neutron optical components that have been inserted in the setup, mainly because of the transmission $T\sim 0.4$ of the supermirror for the $+z$-spin component (and $T=0$ for $-z$).


\end{document}